\documentclass[twocolumn,showpacs,amsmath,amssymb]{revtex4}

\usepackage{graphicx}% Include figure files
\usepackage{dcolumn}% Align table columns on decimal point
\usepackage{bm}% bold math
\usepackage{amssymb}
\usepackage{amsmath}

\usepackage{epsf}
\usepackage{subfigure}
\usepackage{epstopdf}
\newcommand{\bra}[1]{\langle #1|}
\newcommand{\ket}[1]{|#1\rangle}

\newcommand{\mean}[1]{\langle #1 \rangle}
\newcommand{\trace}{{\rm tr}}

\newcommand{\sbra}[1]{\langle\!\langle #1|}
\newcommand{\sket}[1]{|#1\rangle\!\rangle}
\newcommand{\sbraket}[2]{\langle\!\langle #1|#2\rangle\!\rangle}

\newcommand{\e}{{\rm e}}

\begin{document}

\title{Quantum master equation for the microcanonical ensemble}

\author{Massimiliano Esposito}
%\email{mesposit@ulb.ac.be}
\author{Pierre Gaspard}
%\email{gaspard@ulb.ac.be}
\affiliation{Center for Nonlinear Phenomena and Complex Systems,\\
Universit{\'e} Libre de Bruxelles, Code Postal 231, Campus Plaine,
B-1050 Brussels, Belgium.}

\date{\today}

\begin{abstract}
By using projection superoperators, we present a new derivation
of the quantum master equation first obtained by the Authors in
Phys. Rev. E {\bf 68}, 066112 (2003).
We show that this equation describes the dynamics of a subsystem
weakly interacting with an environment of finite heat capacity
and initially described by a microcanonical distribution.
After applying the rotating wave approximation to the
equation, we show that the subsystem dynamics preserves the
energy of the total system (subsystem plus environment) and
tends towards an equilibrium state which corresponds to
equipartition inside the energy shell of the total system.
For infinite heat capacity environments, this equation reduces
to the Redfield master equation for a subsystem interacting
with a thermostat.
These results should be of particular interest to describe
relaxation and decoherence in nanosystems where the environment
can have a finite number of degrees of freedom and the equivalence
between the microcanonical and the canonical ensembles is thus not
always guaranteed.
\end{abstract}



\maketitle

%%%%%%%%%%%%%%%%%%%%%%%%%%%%%%%%%%%%%%%%%%%%%%%%%%%%%%%%%%%%%%%%%%%%
\section{Introduction \label{intro}}

Irreversible processes do not occur in isolated quantum
systems with a finite number of quantum levels.
In order to understand relaxation toward equilibrium
in these systems, one needs to take into account the
effect of their interaction with their environment.
The usual way to proceed is to consider a total Hamiltonian
system composed of a subsystem coupled to an environment.
The subsystem dynamics is then described by the reduced
density matrix of the subsystem which is obtained by
tracing out the degrees of freedom of the environment
from the density matrix of the total system.
In order to drive the subsystem into an effective irreversible
process over reasonably long-time scales (of the order of the
Heisenberg time scale of the environment) on need to assume
a quasi-continuous environment spectrum.
This is typically valid when the energy spacing between
the unperturbed quantum levels of the total system which
are coupled together by the subsystem-environment interaction
is sufficiently small to make the interaction between these
levels effective enough to "mix" them \cite{EspositoPRE03b}.
Once this condition is satisfied, the generic way to obtain a
closed master equation for the reduced density matrix of the
subsystem is to use perturbation theory and the Born-Markov
approximation which implicitly rely on the assumption that
the environment has an infinite heat capacity and cannot be
affected by the system dynamics.
The environment thus plays the role of a thermostat for
the subsystem and the descriptions in the canonical and
microcanonical ensembles are equivalent \cite{KuboB98a}.
The resulting quantum master equation is called the Redfield
equation \cite{Red57,KampenB97,KuboB98b,GaspNaga99,Breuer02}.
Since the Redfield equation can break the positivity of the
subsystem density matrix it is sometimes simplified further
by time averaging the equation in the interaction representation
(rotating wave approximation) in order to get a master equation
of Lindblad form which preserves positivity
\cite{Haake73,Spohn78,Spohn80,CohenTann96,Gardiner00}.

In the present paper, we consider situations where the environment
has a quasi-continuous spectrum but a finite heat capacity.
This means that the energy quanta of the subsystem may
significantly affect the microcanonical temperature of the
environment so that the equivalence between the microcanonical
and canonical statistical ensembles is compromised and only
the microcanonical ensemble can be used {\it a priori}.
Such situations should be generic since the density of state of
a system growth exponentially with the number of degrees of
freedom whereas the heat capacity only growth linearly.
This means that there exists a range in the number of degrees of
freedom where the quasi-continuous assumption can be satisfied
without necessarily implying an infinite heat capacity.
This domain where kinetic processes can occur in the subsystem
but for which the usual master equations fail should be of
particular importance in nanosystems where the number of degrees
of freedom constituting the environment is not always large
enough to be supposed infinite.
In this sense, the present work can be viewed as a contribution to
the recent attempts to apply statistical physics to small systems
\cite{Jensen85,Gross00,Jellinek00,Shirts02,Gross05,Janke06}.
Our results should also be of relevance for systems which
display negative heat capacities in the thermodynamic limit and 
to which the microcanonical ensemble description applies.
A well known example is provided by coupled 
spin systems because their spectra remain bounded and forms 
bands of finite extension
\cite{Barre01,Cohen01,Shepelyansky01,Barre02}.
Other known examples of such systems are reported in Refs.
\cite{Thirring71,Gross97,Ruffo02,Touchette04}.

The plan of the paper is the following.
In Section \ref{newder}, we give a new and more enlightening
derivation of the equation of Ref. \cite{EspoGaspPRE03a}
by using projection superoperators.
In Section \ref{RWA}, we apply the rotating wave approximation
to our equation and prove that it conserves the energy of the
total system.
We also demonstrate that the subsystem relaxes to an
equilibrium distribution which corresponds to equipartition
inside the energy shell of the total system as expected in the
microcanonical ensemble.
In Section \ref{equivens}, we show that, for large heat capacity,
our equation reduces to the Redfield equation and we discuss in
detail the conditions under which the master equations for infinite
thermostat fail to describe the correct subsystem dynamics and need
to be replaced by our master equation.
Conclusions are drawn in Section \ref{conclusion}.

%%%%%%%%%%%%%%%%%%%%%%%%%%%%%%%%%%%%%%%%%%%%%%%%%%%%%%%%%%%%%%%%%%%%
\section{New derivation}\label{newder}

In this section, with the help of projection superoperators, we give
a different derivation of the quantum master equation first
derived in Ref. \cite{EspoGaspPRE03a}.

We consider the dynamics of a system with a Hamiltonian of the form
\begin{eqnarray}
H = H_{0} + \lambda V \;,
\label{A1111a}
\end{eqnarray}
where $\lambda$ is a small dimensionless parameter.
The density
matrix $\rho(t)$ of this total system
obeys the von Neumann equation
\begin{eqnarray}
\dot{\rho}(t) &=& {\cal L} \rho(t)
= -\frac{i}{\hbar} [H,\rho (t)] \label{A1111b} \\
&=& ({\cal L}_0 + \lambda {\cal L}') \rho(t)
=-\frac{i}{\hbar} [H_0,\rho(t)]
- \lambda \frac{i}{\hbar} [V,\rho(t)] \nonumber \; .
\end{eqnarray}
The solution of the von Neumann equation reads
\begin{eqnarray}
\rho (t) = {\cal U}(t) \rho (0) = \e^{{\cal L}t} \rho (0)
\label{A1111c}
\end{eqnarray}
In the interaction representation where
\begin{eqnarray}
\rho_I(t) &=& \e^{-{\cal L}_0 t} \rho(t)
= \e^{\frac{i}{\hbar} H_0 t} \rho(t) \e^{-\frac{i}{\hbar} H_0 t} \; , \\
{\cal L}'_I(t) &=& \e^{-{\cal L}_0 t} {\cal L}'(t) \e^{{\cal L}_0 t} \; ,
\label{A1111d}
\end{eqnarray}
the von Neumann equation takes the simple form
\begin{eqnarray}
\dot{\rho}_I(t) &=& \lambda {\cal L}'_I(t) \rho_I(t)
= -\lambda\frac{i}{\hbar} \lbrack V_I(t),\rho_I(t)\rbrack \;.
\label{A1111e}
\end{eqnarray}
By integrating Eq. (\ref{A1111e}) and truncating it to order $\lambda^2$,
we get the perturbative expansion
\begin{eqnarray}
\rho_I(t)&=&{\cal W}(t) \rho(0)
=\e^{-{\cal L}_0 t} \e^{{\cal L} t} \rho(0) \label{A1111f} \\
&=&\left[ {\cal W}_0(t) + \lambda {\cal W}_1(t)
+ \lambda^2 {\cal W}_2(t) + {\cal O}(\lambda^3) \right] \rho(0)
\; , \nonumber
\end{eqnarray}
where
\begin{eqnarray}
{\cal W}_0(t) &=& I \; ; \label{A1111g}\\
\; {\cal
W}_1(t)&=&\int_{0}^{t} dT \, {\cal L}'_I(T) \; ; \nonumber\\
{\cal W}_2(t)&=&\int_{0}^{t}dT \int_{0}^{T} d\tau \, {\cal L}'_I(T)
{\cal L}'_I(T-\tau)\; . \nonumber
\end{eqnarray}
The inverse of ${\cal W}(t)$ reads
\begin{eqnarray}
{\cal W}^{-1}(t)&=&{\cal W}_0(t) - \lambda {\cal W}_1(t)
\label{A1111h}\\
&&+ \lambda^2 \left[{\cal W}_1^2(t)-{\cal W}_2(t) \right]
+ {\cal O}(\lambda^3) \; . \nonumber
\end{eqnarray}
Indeed, one can check that ${\cal W}(t){\cal W}^{-1}(t)
= 1 + {\cal O}(\lambda^3)$.
For later purpose, we also notice that
\begin{eqnarray}
\dot{{\cal W}}(t) {\cal A} {\cal W}^{-1}(t)&=&
\lambda \dot{{\cal W}}_1(t) {\cal A} \label{A1111i} \\
&&\hspace{-1cm}+ \lambda^2 \left[\dot{{\cal W}}_2(t) {\cal A}
- \dot{{\cal W}}_1(t) {\cal A} {\cal W}_{1}(t) \right]
+ {\cal O}(\lambda^3) \; . \nonumber
\end{eqnarray}

Now, we consider a subsystem $S$ interacting with its
environment $B$. The Hamiltonian of this system is given
by Eq. (\ref{A1111a}) where
\begin{eqnarray}
H_0 = H_S + H_B \; ; \; V = \sum_{\kappa} S^{\kappa} B_{\kappa}
\; . \label{A1111j}
\end{eqnarray}
$S^{\kappa}$ (resp. $B_{\kappa}$) is a coupling operator of system
$S$ (resp. $B$).
We will use the index $s$ (respectively $b$) to label the eigenstates
of the Hamiltonian of system $S$ (resp. $B$).

We will use Liouville space where operators are mapped into
vectors  and superoperators into matrices \cite{Mukamel95}.
We recall some basic definitions
\begin{eqnarray}
&&{\rm scalar \; product:} \ \ \sbraket{A}{B} \equiv \trace A^{\dagger} B \; ,
\label{A1111k}\\
&&{\rm identity:} \ \ {\cal I} \equiv \sum_{n,n'} \sket{nn'} \sbra{nn'} \; ,
\label{A1111l}\\
&&\hspace*{0cm} \sket{nn'} \leftrightarrow \ket{n} \bra{n'} \; , \;
\sbra{nn'} \leftrightarrow \ket{n'} \bra{n} \; .
\label{A1111m}
\end{eqnarray}
Useful consequences of these definitions are
\begin{eqnarray}
\sbraket{nn'}{\bar{n}\bar{n}'}&=&\delta_{n\bar{n}} \delta_{n'\bar{n}'}
\label{A1111n}\\
\sbraket{nn'}{A}&=&\bra{n} A \ket{n'} \; . \label{A1111o}
\end{eqnarray}
An operator $A$ in the total space which can be written as a product of
a system and reservoir operator $A=A_S \otimes A_B$ will read in
Liouville space $\sket{A} = \sket{A_S} \otimes \sket{A_B}$.\\
We now define the following projection superoperators
which act on the Liouville space of system $B$
\begin{eqnarray}
{\cal P} &=& \sum_{b} \sket{bb} \sbra{bb} \label{A1111p}  \\
{\cal Q} &=& \sum_{b,b'} \sket{bb'} \sbra{bb'} (1-\delta_{bb'})
\label{A1111p} \; .
\end{eqnarray}
and which satisfy the usual properties of projection
superoperators ${\cal P} + {\cal Q} = {\cal I}_B, {\cal P}^2 =
{\cal P}, {\cal Q}^2 = Q$ and ${\cal P}{\cal Q} =
{\cal Q}{\cal P} = 0$.
${\cal P}$, when acting on the density matrix of the total system,
eliminates the environment coherences but keeps the environment
populations and leaves unaffected the subsystem degrees of freedom.
Similar projection superoperators have been recently used in Refs.
\cite{Espositothese,Budini,BreuerGemmer,Breuer}.
For comparison, the projection superoperator used
to derive the Redfield equation reads
\begin{eqnarray}
{\cal P}_{Red} &=& \sum_{b} \sket{\rho^{eq}} \sbra{bb}
\label{A1111pred} \; ,
\end{eqnarray}
where $\rho^{eq}$ is the equilibrium density matrix of the environment.
The two projection superoperators act differently 
on the density matrix $\rho(t)$.
In the Liouville space of the total system, we get
respectively
\begin{eqnarray}
{\cal P} \sket{\rho(t)} &=& \sum_b \sbraket{bb}{\rho(t)} \otimes \sket{bb}
\label{A1111pbisA}\\
{\cal P}_{Red} \sket{\rho(t)} &=& \sket{\rho_S(t)} \otimes  \sket{\rho^{eq}}
\label{A1111pbisB} \; .
\end{eqnarray}
$\sbraket{bb}{\rho(t)}$ is a density matrix in the system
space which depends on the environment state $b$.
${\cal P}$ therefore correlates the system state with the environment state.
On the contrary, ${\cal P}_{Red}$ assumes that the system reduced density
matrix $\rho_S(t) \equiv \sum_b \sbraket{bb}{\rho(t)} = \trace_B \rho(t)$
is independent from the environment state which always remain at equilibrium.\\
We now let ${\cal P}$ and ${\cal Q}$ act on the density matrix of
the total system in the interaction form (\ref{A1111f}) and find
\begin{eqnarray}
{\cal P} \sket{\rho_I(t)} &=& {\cal P} {\cal W}(t)
({\cal P}+{\cal Q}) \sket{\rho_I(0)} \label{A1111qa} \\
{\cal Q} \sket{\rho_I(t)} &=& {\cal Q} {\cal W}(t)
({\cal P}+{\cal Q}) \sket{\rho_I(0)} \label{A1111qb} \; .
\end{eqnarray}
 From now on, we will consider initial conditions such that
${\cal Q} \sket{\rho(0)}=0$. This means that the environment part
of the initial condition is diagonal in the environment eigenbasis
and is thus invariant under the evolution when $\lambda = 0$.
Taking the time derivative of Eq. (\ref{A1111qa}) and Eq. (\ref{A1111qb})
and using $\sket{\rho_I(0)}={\cal W}^{-1}(t) \sket{\rho_I(t)}$, we get
\begin{eqnarray}
{\cal P} \sket{\dot{\rho}_I(t)} &=&
{\cal P} \dot{{\cal W}}(t) {\cal P} {\cal W}^{-1}(t)
{\cal P} \sket{\rho_I(t)} \label{A1111ra} \\
&&+{\cal P} \dot{{\cal W}}(t) {\cal P} {\cal W}^{-1}(t)
{\cal Q} \sket{\rho_I(t)} \nonumber \\
{\cal Q} \sket{\dot{\rho}_I(t)} &=&
{\cal Q} \dot{{\cal W}}(t) {\cal P} {\cal W}^{-1}(t)
{\cal P} \sket{\rho_I(t)} \label{A1111rb} \\
&&+{\cal Q} \dot{{\cal W}}(t) {\cal P} {\cal W}^{-1}(t)
{\cal Q} \sket{\rho_I(t)} \; .\nonumber
\end{eqnarray}
These equations are still exact. If we restrict ourselves
to second-order perturbation theory, we can obtain the
important result that the ${\cal P}$ projected density
matrix evolution is decoupled from the ${\cal Q}$ projected part.
Indeed, with the help of Eq. (\ref{A1111i}), we have
\begin{eqnarray}
{\cal P} \dot{{\cal W}}(t) {\cal P} {\cal W}^{-1}(t) {\cal Q}
&=& \lambda {\cal P} \dot{{\cal W}}_1(t) {\cal P} {\cal Q}
\label{A1111s} \\
&&\hspace{-2.3cm} +\lambda^2 {\cal P} \dot{{\cal W}}_2(t)
{\cal P} {\cal Q} - \lambda^2 {\cal P} \dot{{\cal W}}_1(t)
{\cal P} {\cal W}_1(t) {\cal Q} + {\cal O}(\lambda^3) \; .
\nonumber
\end{eqnarray}
The two first terms of the right-hand side are zero because
${\cal P} {\cal Q} = 0$ and the third one also because
\begin{eqnarray}
{\cal P} \dot{{\cal W}}_1(t) {\cal P} &=& \sum_{b,b'} \sket{bb}
\sbra{bb} {\cal L}_I'(t) \sket{b'b'} \sbra{b'b'} \label{A1111t}\\
&&\hspace{-2cm}=-\frac{i}{\hbar} \sum_{\kappa}
S^{\kappa}(t) \sum_{b,b'} \sket{bb}
\bra{b} \left[ B_{\kappa}(t),\ket{b'}\bra{b'} \right] \ket{b}
\sbra{b'b'}\; , \nonumber
\end{eqnarray}
where $\bra{b} \left[ B_{\kappa}(t),\ket{b'}\bra{b'}
\right] \ket{b}=0$.

After having showed that the relevant projected density
matrix evolves in an autonomous way, we will now evaluate
the generator of its evolution using second-order
perturbation theory. Again using Eq. (\ref{A1111i}), we find that
\begin{eqnarray}
{\cal P} \dot{{\cal W}}(t) {\cal P} {\cal W}^{-1}(t) {\cal P}
&=& \lambda {\cal P} \dot{{\cal W}}_1(t) {\cal P}
\label{A1111u} \\
&&\hspace{-2.3cm} +\lambda^2 {\cal P} \dot{{\cal W}}_2(t)
{\cal P} - \lambda^2 {\cal P} \dot{{\cal W}}_1(t)
{\cal P} {\cal W}_1(t) {\cal P} + {\cal O}(\lambda^3) \; .
\nonumber
\end{eqnarray}
The only term of right-hand side which is not zero is the
second one [see Eq. (\ref{A1111t})] whereupon we get
\begin{eqnarray}
{\cal P} \sket{\dot{\rho}_I(t)} &=& \lambda^2 {\cal P}
\int_{0}^{t} d\tau {\cal L}'_I(t) {\cal L}'_I(t-\tau)
{\cal P} \sket{\rho_I(t)} \nonumber \\
&&+ {\cal O}(\lambda^3) \; .\label{A1111v}
\end{eqnarray}
Now leaving the interaction representation and using the
fact that ${\cal P} \e^{-{\cal L}_0 t} = \e^{-{\cal L}_S t}
{\cal P}$, we obtain
\begin{eqnarray}
{\cal P} \sket{\dot{\rho}(t)} &=&
{\cal L}_S {\cal P} \sket{\rho(t)} \label{A1111w} \\
&&\hspace{-2cm}+ \lambda^2 \e^{{\cal L}_S t} {\cal P} \int_{0}^{t}
d\tau {\cal L}'_I(t) {\cal L}'_I(t-\tau) \e^{-{\cal L}_S t}
{\cal P} \sket{\rho(t)} \nonumber + {\cal O}(\lambda^3) \; .\nonumber
\end{eqnarray}

Now, we define the quantity $P(E_b,t)$ that will become the
fundamental quantity of our theory
\begin{eqnarray}
\frac{P(E_b,t)}{n(E_b)} \equiv
\sbra{bb} {\cal P} \sket{\rho(t)} \label{A1111x} \; ,
\end{eqnarray}
where $n(E_b) = \trace_B \delta(E_b-H_B)$ is the density of state
of the environment. Equation (\ref{A1111w}) can thus be rewritten as
\begin{eqnarray}
\frac{\dot{P}(E_b,t)}{n(E_b)} &=& {\cal L}_S \frac{P(E_b,t)}{n(E_b)}
+ \lambda^2 \sum_{b'} \int_{0}^{t} d\tau \label{A1111y} \\
&&\hspace{-1cm}  \times \; \e^{{\cal L}_S t} \sbra{bb} {\cal L}_I'(t)
{\cal L}_I'(t-\tau)
\sket{b'b'} \e^{-{\cal L}_S t} \frac{P(E_{b'},t)}{n(E_{b'})} \; ,
\nonumber
\end{eqnarray}
where we stop explicitly writing $+{\cal O}(\lambda^3)$ from now on.
Equation (\ref{A1111y}) can be evaluated further and we get
\begin{eqnarray}
\frac{\dot{P}(E_b,t)}{n(E_b)} &=& {\cal L}_S \frac{P(E_b,t)}{n(E_b)}
- \frac{\lambda^2}{\hbar^2} \sum_{\kappa \kappa'} \sum_{b'} \int_{0}^{t} d\tau
\label{A1111z} \\
&&\hspace{-1.4cm}  \e^{-\frac{i}{\hbar} H_S t} \bra{b}
\Big[ S^{\kappa}(t) B_{\kappa}(t), \Big[ S^{\kappa'}(t-\tau)
B_{\kappa'}(t-\tau), \nonumber  \\
&&  \e^{\frac{i}{\hbar} H_S t}
\frac{P(E_{b'},t)}{n(E_{b'})} \e^{-\frac{i}{\hbar} H_S t}
\ket{b'} \bra{b'} \Big] \Big] \ket{b} \; \e^{\frac{i}{\hbar}
H_S t} \; . \nonumber
\end{eqnarray}
By evaluating the commutators, we get the non-Markovian version
of our master equation for environments with discrete spectra
\begin{eqnarray}
\frac{\dot{P}(E_b,t)}{n(E_b)} &=& {\cal L}_S \frac{P(E_b,t)}{n(E_b)}
- \frac{\lambda^2}{\hbar^2} \sum_{\kappa \kappa'} \sum_{b'}
\int_{0}^{t} d\tau \Big\lbrace
\label{A1112a} \\
&&\hspace{-1cm}+\e^{i \omega_{bb'} \tau} \bra{b} B_{\kappa} \ket{b'}
\bra{b'} B_{\kappa'} \ket{b} S^{\kappa} S^{\kappa'}(-\tau)
\frac{P(E_{b},t)}{n(E_{b})} \nonumber \\
&&\hspace{-1cm}-\e^{i \omega_{bb'} \tau} \bra{b} B_{\kappa} \ket{b'}
\bra{b'} B_{\kappa'} \ket{b} S^{\kappa}
\frac{P(E_{b'},t)}{n(E_{b'})} S^{\kappa'}(-\tau) \nonumber \\
&&\hspace{-1cm}-\e^{-i \omega_{bb'} \tau} \bra{b} B_{\kappa'} \ket{b'}
\bra{b'} B_{\kappa} \ket{b} S^{\kappa'}(-\tau)
\frac{P(E_{b'},t)}{n(E_{b'})} S^{\kappa} \nonumber \\
&&\hspace{-1cm}+\e^{-i \omega_{bb'} \tau} \bra{b} B_{\kappa'} \ket{b'}
\bra{b'} B_{\kappa} \ket{b} \frac{P(E_{b},t)}{n(E_{b})}
S^{\kappa'}(-\tau) S^{\kappa}  \Big\rbrace \; , \nonumber
\end{eqnarray}
where $\omega_{bb'}=(E_b-E_{b'})/\hbar$ are the Bohr frequencies of
the environment.\\
We assume now that the spectra of the environment is dense enough
to be treated as quasicontinuum so that we can use the following
equivalences
\begin{eqnarray}
&&\hspace{-0.35cm}P(E_b,t) = n(E_b) \trace_B \ket{b}
\bra{b} \rho(t) \label{A1112b}\\&&\hspace{3cm}\to P(\epsilon,t)
= \trace_B \delta(\epsilon-H_B) \rho(t) \; , \nonumber\\
&&\hspace{-0.35cm}\sum_{b'} \to \int d\epsilon' n(\epsilon') \;, \ \
n(E_b) \to n(\epsilon) \; , \ \ \bra{b} B \ket{b'} \to
\bra{\epsilon} B \ket{\epsilon'} \; . \nonumber
\end{eqnarray}
The non-Markovian version of our master equation for environments
with a quasi-continuous spectrum is therefore
\begin{eqnarray}
\dot{P}(\epsilon,t) &=& {\cal L}_S P(\epsilon,t)
- \frac{\lambda^2}{\hbar^2} \sum_{\kappa \kappa'} \int d\epsilon'
\int_{0}^{t} d\tau \Big\lbrace \label{A1112c} \\
&&\hspace{-1.5cm}+\e^{i (\epsilon-\epsilon') \tau/\hbar} n(\epsilon')
\bra{\epsilon} B_{\kappa} \ket{\epsilon'}
\bra{\epsilon'} B_{\kappa'} \ket{\epsilon}
S^{\kappa} S^{\kappa'}(-\tau) P(\epsilon,t) \nonumber \\
&&\hspace{-1.5cm}-\e^{i (\epsilon-\epsilon') \tau/\hbar} n(\epsilon)
\bra{\epsilon} B_{\kappa} \ket{\epsilon'}
\bra{\epsilon'} B_{\kappa'} \ket{\epsilon} S^{\kappa}
P(\epsilon',t) S^{\kappa'}(-\tau) \nonumber \\
&&\hspace{-1.5cm}-\e^{-i (\epsilon-\epsilon') \tau/\hbar} n(\epsilon)
\bra{\epsilon} B_{\kappa'} \ket{\epsilon'}
\bra{\epsilon'} B_{\kappa} \ket{\epsilon} S^{\kappa'}(-\tau)
P(\epsilon',t) S^{\kappa} \nonumber \\
&&\hspace{-1.5cm}+\e^{-i (\epsilon-\epsilon') \tau/\hbar} n(\epsilon')
\bra{\epsilon} B_{\kappa'} \ket{\epsilon'}
\bra{\epsilon'} B_{\kappa} \ket{\epsilon} P(\epsilon,t)
S^{\kappa'}(-\tau) S^{\kappa} \Big\rbrace \; . \nonumber
\end{eqnarray}
This equation was first obtained in Ref. \cite{EspoGaspPRE03a}.
Notice that the reduced density matrix of the subsystem is obtained
from the quantity $P(\epsilon,t)$ using
\begin{eqnarray}
\rho_S(t) = \trace_B \rho(t) = \int d\epsilon P(\epsilon,t)
\label{A1112d} \; .
\end{eqnarray}
The quantity $P(\epsilon,t)$ can be seen as an environment
energy distributed subsystem density matrix. One should also
realize that Eq. (\ref{A1112c}) is a closed equation for the
$P(\epsilon,t)$'s but cannot be converted without approximations
into a closed equation for $\rho_S(t)$ using (\ref{A1112d}).
This is a first indication that Eq. (\ref{A1112c}) contains more
information than what can be contained in a closed equation
for $\rho_S(t)$.\\

In the environment space, the equilibrium correlation function
between two coupling operators $B_{\kappa}$ and $B_{\kappa'}$ is
\begin{eqnarray}
\alpha_{\kappa \kappa'}(t) = \trace_B \rho^{eq} \e^{\frac{i}{\hbar}
H_B t} B_{\kappa}
\e^{-\frac{i}{\hbar} H_B t} B_{\kappa'} \; , \label{A1112e}
\end{eqnarray}
where $\rho^{eq}$ is the equilibrium density matrix of the environment.
If the environment is described by a microcanonical distribution at
energy $\epsilon$
\begin{eqnarray}
\rho^{eq}_{mic}(\epsilon) = \delta(\epsilon-H_B)/n(\epsilon) \;,
\label{A1112f}
\end{eqnarray}
the correlation function reads
\begin{eqnarray}
\alpha_{\kappa \kappa'}(\epsilon,t)=\int d\epsilon' n(\epsilon')
\e^{i(\epsilon-\epsilon')t/\hbar} \bra{\epsilon} B_{\kappa} \ket{\epsilon'}
\bra{\epsilon'} B_{\kappa'} \ket{\epsilon} \; . \label{A1112g}
\end{eqnarray}
Its Fourier transform [$\tilde{\alpha}(\omega)=\int_{-\infty}^{\infty}
\frac{dt}{2\pi} \e^{i \omega t} \alpha(t)$] is
\begin{eqnarray}
\tilde{\alpha}_{\kappa\kappa'}(\epsilon,\omega) =
\hbar n(\epsilon+\hbar \omega) \bra{\epsilon} B_{\kappa}
\ket{\epsilon+\hbar \omega} \bra{\epsilon+\hbar \omega}
B_{\kappa'} \ket{\epsilon} \; . \label{A1112h}
\end{eqnarray}
This quantity has a useful physical interpretation.
$\sum_{\kappa \kappa'} \tilde{\alpha}_{\kappa \kappa'}$ is, up to a
factor $\hbar^2/(\pi \lambda^2)$, the Fermi golden rule transition
rate for the environment in a microcanonical distribution at energy
$\epsilon$ to absorb (resp. emit) a quantum of energy $\hbar \omega$
when submitted to a periodic perturbation
$\sum_{\kappa} B_{\kappa} \cos \omega t$.
Using (\ref{A1112h}), we can easily verify the important property
\begin{eqnarray}
\tilde{\alpha}_{\kappa\kappa'}(\epsilon,\omega) n(\epsilon)
= \tilde{\alpha}_{\kappa'\kappa}(\epsilon+\hbar \omega,-\omega)
n(\epsilon+\hbar \omega) \; . \label{A1112i}
\end{eqnarray}
We can now rewrite Eq. (\ref{A1112c}) as
\begin{eqnarray}
\dot{P}(\epsilon,t) &=& {\cal L}_S P(\epsilon,t)
+ \frac{\lambda^2}{\hbar^2} \sum_{\kappa \kappa'}
\int_{0}^{t} d\tau \int_{-\infty}^{\infty} d\omega
\Big\lbrace \label{A1112j} \\
&&\hspace{-1.5cm}-\e^{-i \omega \tau}
\tilde{\alpha}_{\kappa\kappa'}(\epsilon,\omega)
S^{\kappa} S^{\kappa'}(-\tau) P(\epsilon,t) \nonumber \\
&&\hspace{-1.5cm}-\e^{i \omega \tau}
\tilde{\alpha}_{\kappa'\kappa}(\epsilon,\omega)
P(\epsilon,t) S^{\kappa'}(-\tau) S^{\kappa} \nonumber \\
&&\hspace{-1.5cm}+\e^{-i \omega \tau}
\tilde{\alpha}_{\kappa'\kappa}(\epsilon+\hbar \omega,-\omega)
S^{\kappa} P(\epsilon+\hbar \omega,t) S^{\kappa'}(-\tau) \nonumber \\
&&\hspace{-1.5cm}+\e^{i \omega \tau}
\tilde{\alpha}_{\kappa\kappa'}(\epsilon+\hbar \omega,-\omega)
S^{\kappa'}(-\tau) P(\epsilon+\hbar \omega,t) S^{\kappa}
\Big\rbrace \; .  \nonumber
\end{eqnarray}
This equation is the same non-Markovian master equation as
(\ref{A1112c}) but written in a more compact and intuitive
form. It is explicit now that the effect of the environment
on the subsystem dynamics only enters the description
via the environment microcanonical correlation function.
In standard master equations, the same is true but with
canonical instead of microcanonical correlation functions.
The Markovian version of this equation is obtained by
taking the upper bound of the integral to infinity
$\int_{0}^{t} \to \int_{0}^{\infty}$.
This approximation is well known in the literature and is
justified when the environment correlation function decays
on time scales $\tau_c$ much shorter then the
fastest time scale of the free subsystem evolution $\tau_S$
(typically given by the inverse of the largest subsystem Bohr
frequency).

A remark concerning the terminology is required at this point.
In our terminology, the generator of time evolution is 
said non-Markovian if it explicitly depends on time whether 
or not it acts on $\rho_S$ or $P(\epsilon)$.
An alternative definition (see \cite{Breuer}) defines the
system dynamics as non-Markovian if it cannot be described
by a time-independent generator acting on $\rho_S$.
With this definition, Eq. (\ref{A1112j}) with the upper 
bound of the time integral taken to infinity 
would describe a non-Markovian dynamics.

%%%%%%%%%%%%%%%%%%%%%%%%%%%%%%%%%%%%%%%%%%%%%%%%%%%%%%%%%%%%%%%%%%%%
\section{Rotating wave approximation}\label{RWA}

In this section, we will use the rotating wave approximation
(RWA) which consist in time averaging the Markovian version
of the master equation in the interaction form.
When applying this approximation to Eq. (\ref{A1112j}), the
equation takes a simple form which allows to prove important results.\\

The RWA is most commonly used in quantum optic
\cite{Breuer02,CohenTann96,Gardiner00} because the free subsystem
dynamics generally evolves on times scales $\tau_S$ which are much
faster than the relaxation time scales $\tau_r$ induced by the coupling
to the environment.
The master equation in the interacting representation evolves then very
slowly compared to the Bohr frequencies of the subsystem which can
therefore be averaged out.
In other words, the RWA is justified if the time scale separation
$\tau_c \ll \tau_S \ll \tau_r$ exist.
In the mathematical physics, peoples usually refer to this averaging
procedure (which is performed after a rescaling of time $t'= \lambda^2 t$)
as the weak coupling limit \cite{Spohn78,Spohn80} since the smaller
the coupling the longer $\tau_r$.
Their motivation is to impose a Lindblad form to the master equation
generator in order to ensure the positivity of the subsystem density
matrix \cite{Breuer02,Spohn80,Lindblad76}.
The same situation occurs in our case.
In Ref. \cite{Breuer}, Breuer has generalized the Lindblad theory
to generators which act on projected total density matrices of the
type (\ref{A1111pbisA}) which correlate the system-reservoir dynamics.
Eq. (\ref{A1112j}) is not of the generalized Lindblad form and could
in principle lead to positivity breakdown similarly as the Redfield
equation \cite{GaspNaga99,Silbey92,Silbey05,Tannor97}.
The RWA can be used to guaranty that our equation preserves
the positivity of the subsystem density matrix.
By writing Eq. (\ref{A1112j}) in the interaction representation and by
projecting the resulting equation in the subsystem eigenbasis, we get
\begin{eqnarray}
\dot{P}^I_{ss'}(\epsilon,t) &=&
\frac{\lambda^2}{\hbar^2} \sum_{\kappa \kappa'} \sum_{\bar{s}\bar{s}'}
\int_{0}^{\infty} d\tau \int_{-\infty}^{\infty} d\omega \Big\lbrace
\label{B11111} \\
&&\hspace{-1.5cm}-\e^{-i (\omega+\omega_{\bar{s}\bar{s}'}) \tau}
\e^{i \omega_{s\bar{s}'} t} \tilde{\alpha}_{\kappa\kappa'}(\epsilon,\omega)
S^{\kappa}_{s\bar{s}} S^{\kappa'}_{\bar{s}\bar{s}'}
P^{I}_{\bar{s}'s'}(\epsilon,t) \nonumber \\
&&\hspace{-1.5cm}-\e^{i (\omega-\omega_{\bar{s}\bar{s}'}) \tau}
\e^{i \omega_{\bar{s}s'} t} \tilde{\alpha}_{\kappa'\kappa}(\epsilon,\omega)
P^I_{s\bar{s}}(\epsilon,t) S^{\kappa'}_{\bar{s}\bar{s}'}
S^{\kappa}_{\bar{s}'s'}
\nonumber \\
&&\hspace{-1.5cm}+\e^{-i (\omega+\omega_{\bar{s}'s'}) \tau}
\e^{i (\omega_{s\bar{s}}+\omega_{\bar{s}'s'}) t} \nonumber \\
&&\hspace{-0.cm}\tilde{\alpha}_{\kappa'\kappa}(\epsilon+\hbar \omega,-\omega)
S^{\kappa}_{s\bar{s}} P^{I}_{\bar{s}\bar{s}'}(\epsilon+\hbar \omega,t)
S^{\kappa'}_{\bar{s}'s'} \nonumber \\
&&\hspace{-1.5cm}+\e^{i (\omega-\omega_{s\bar{s}}) \tau}
\e^{i (\omega_{s\bar{s}}+\omega_{\bar{s}'s'}) t} \nonumber \\
&&\hspace{-0.cm}\tilde{\alpha}_{\kappa\kappa'}(\epsilon+\hbar \omega,-\omega)
S^{\kappa'}_{s\bar{s}} P^{I}_{\bar{s}\bar{s}'}(\epsilon+\hbar \omega,t)
S^{\kappa}_{\bar{s}'s'} \Big\rbrace \; . \nonumber
\end{eqnarray}
The RWA consist in time averaging
$\lim_{T\to\infty}\frac{1}{2T}\int_{-T}^{T}dt$
the right hand side of Eq. (\ref{B11111}) so that
\begin{eqnarray}
\dot{P}^I_{ss'}(\epsilon,t) &=&
\frac{\lambda^2}{\hbar^2} \sum_{\kappa \kappa'} \sum_{\bar{s}\bar{s}'}
\int_{0}^{\infty} d\tau \int_{-\infty}^{\infty} d\omega \Big\lbrace
\label{B11112} \\
&&\hspace{-1.5cm}-\delta_{s\bar{s}'} \e^{-i (\omega+\omega_{\bar{s}s}) \tau}
\tilde{\alpha}_{\kappa\kappa'}(\epsilon,\omega)
S^{\kappa}_{s\bar{s}} S^{\kappa'}_{\bar{s}s}
P^{I}_{ss'}(\epsilon,t) \nonumber \\
&&\hspace{-1.5cm}-\delta_{\bar{s}s'} \e^{i (\omega-\omega_{s'\bar{s}'}) \tau}
\tilde{\alpha}_{\kappa'\kappa}(\epsilon,\omega) P^I_{ss'}(\epsilon,t)
S^{\kappa'}_{s'\bar{s}'} S^{\kappa}_{\bar{s}'s'}
\nonumber \\
&&\hspace{-1.5cm}+\left[(1-\delta_{ss'})\delta_{s\bar{s}}\delta_{\bar{s}'s'}
+\delta_{ss'}\delta_{\bar{s}'\bar{s}}\right]
\e^{-i (\omega+\omega_{\bar{s}'s'}) \tau} \nonumber \\
&&\hspace{-0.cm}\tilde{\alpha}_{\kappa'\kappa}(\epsilon+\hbar \omega,-\omega)
S^{\kappa}_{s\bar{s}} P^{I}_{\bar{s}\bar{s}'}(\epsilon+\hbar \omega,t)
S^{\kappa'}_{\bar{s}'s'} \nonumber \\
&&\hspace{-1.5cm}+\left[(1-\delta_{ss'})\delta_{s\bar{s}}\delta_{\bar{s}'s'}
+\delta_{ss'}\delta_{\bar{s}'\bar{s}}\right]
\e^{i (\omega-\omega_{s\bar{s}}) \tau} \nonumber \\
&&\hspace{-0.cm}\tilde{\alpha}_{\kappa\kappa'}(\epsilon+\hbar \omega,-\omega)
S^{\kappa'}_{s\bar{s}} P^{I}_{\bar{s}\bar{s}'}(\epsilon+\hbar \omega,t)
S^{\kappa}_{\bar{s}'s'} \Big\rbrace \; . \nonumber
\end{eqnarray}
Using $\int_{0}^{\infty} d\tau \e^{\pm i \omega \tau} =
\pi \delta(\omega) \pm i {\rm P} \frac{1}{\omega}$,
%\label{B11112}
we finally get the Markovian version of our master equation in the RWA
\begin{eqnarray}
\dot{P}^I_{ss'}(\epsilon,t) &=&
\frac{\lambda^2}{\hbar^2} \sum_{\kappa \kappa'} \sum_{\bar{s}\bar{s}'}
\Big\lbrace \label{B11113} \\
&&\hspace{-1.7cm}-\pi \delta_{s\bar{s}'}
\tilde{\alpha}_{\kappa\kappa'}(\epsilon,-\omega_{\bar{s}s})
S^{\kappa}_{s\bar{s}} S^{\kappa'}_{\bar{s}s}
P^{I}_{ss'}(\epsilon,t) \nonumber \\
&&\hspace{-1.7cm}-\pi \delta_{\bar{s}s'}
\tilde{\alpha}_{\kappa'\kappa}(\epsilon,\omega_{s'\bar{s}'})
P^I_{ss'}(\epsilon,t) S^{\kappa'}_{s'\bar{s}'}
S^{\kappa}_{\bar{s}'s'} \nonumber \\
&&\hspace{-1.7cm}+\pi
\left[(1-\delta_{ss'})\delta_{s\bar{s}}\delta_{\bar{s}'s'}
+\delta_{ss'}\delta_{\bar{s}'\bar{s}}\right]
\tilde{\alpha}_{\kappa'\kappa}
(\epsilon-\hbar \omega_{\bar{s}'s'},\omega_{\bar{s}'s'}) \nonumber \\
&&\hspace{2cm} S^{\kappa}_{s\bar{s}}
P^{I}_{\bar{s}\bar{s}'}(\epsilon-\hbar \omega_{\bar{s}'s'},t)
S^{\kappa'}_{\bar{s}'s'} \nonumber \\
&&\hspace{-1.7cm}+\pi
\left[(1-\delta_{ss'})\delta_{s\bar{s}}\delta_{\bar{s}'s'}
+\delta_{ss'}\delta_{\bar{s}'\bar{s}}\right]
\tilde{\alpha}_{\kappa\kappa'}
(\epsilon+\hbar \omega_{s\bar{s}},-\omega_{s\bar{s}})
\nonumber \\ &&\hspace{2cm}
S^{\kappa'}_{s\bar{s}} P^{I}_{\bar{s}\bar{s}'}
(\epsilon+\hbar \omega_{s\bar{s}},t)
S^{\kappa}_{\bar{s}'s'} \nonumber \\
&&\hspace{-1.7cm}+i \delta_{s\bar{s}'}
\int_{-\infty}^{\infty} d\omega {\rm P}
\frac{\tilde{\alpha}_{\kappa\kappa'}(\epsilon,\omega)}
{\omega+\omega_{\bar{s}s}} \;
S^{\kappa}_{s\bar{s}} S^{\kappa'}_{\bar{s}s}
P^{I}_{ss'}(\epsilon,t) \nonumber \\
&&\hspace{-1.7cm}-i \delta_{\bar{s}s'}
\int_{-\infty}^{\infty} d\omega {\rm P}
\frac{\tilde{\alpha}_{\kappa'\kappa}(\epsilon,\omega)}
{\omega-\omega_{s'\bar{s}'}} \;
P^I_{ss'}(\epsilon,t) S^{\kappa'}_{s'\bar{s}'}
S^{\kappa}_{\bar{s}'s'} \nonumber \\
&&\hspace{-1.7cm}-i \left[(1-\delta_{ss'})\delta_{s\bar{s}}\delta_{\bar{s}'s'}
+\delta_{ss'}\delta_{\bar{s}'\bar{s}}\right]
S^{\kappa}_{s\bar{s}} \nonumber \\
&&\hspace{-1.4cm} \int_{-\infty}^{\infty} d\omega {\rm P}
\frac{\tilde{\alpha}_{\kappa'\kappa}(\epsilon+\hbar \omega,-\omega)
P^{I}_{\bar{s}\bar{s}'}(\epsilon+\hbar \omega,t)}
{\omega+\omega_{\bar{s}'s'}} \;
S^{\kappa'}_{\bar{s}'s'} \nonumber \\
&&\hspace{-1.7cm}+i \left[(1-\delta_{ss'})\delta_{s\bar{s}}\delta_{\bar{s}'s'}
+\delta_{ss'}\delta_{\bar{s}'\bar{s}}\right]
S^{\kappa'}_{s\bar{s}} \nonumber \\
&&\hspace{-1.4cm}  \int_{-\infty}^{\infty} d\omega {\rm P}
\frac{\tilde{\alpha}_{\kappa\kappa'}(\epsilon+\hbar \omega,-\omega)
P^{I}_{\bar{s}\bar{s}'}(\epsilon+\hbar \omega,t)}
{\omega-\omega_{s\bar{s}}} \;
S^{\kappa}_{\bar{s}'s'} \Big\rbrace \; . \nonumber
\end{eqnarray}
This equation is a central result of this paper.
It might look a little complicated but is in fact relatively simple.
The first four terms are responsible for the damping of the subsystem
and the four last term are small shifts in the Bohr frequencies of the
subsystem. The populations ($s = s'$) evolve independently from the
coherences ($s \neq s'$) and the coherences evolve independently for
each other following exponentially damped oscillations
\begin{eqnarray}
\dot{P}_{ss'}(\epsilon,t) =
(- \Gamma_{ss'} - i \Omega_{ss'}) P_{ss'}(\epsilon,t)
\label{B11114} \; ,
\end{eqnarray}
where the damping rates are given by
\begin{eqnarray}
\Gamma_{ss'} &=& \frac{\lambda^2}{\hbar^2} \sum_{\kappa \kappa'} \left\{
- 2 \pi \tilde{\alpha}_{\kappa\kappa'}(\epsilon,0)
S^{\kappa'}_{ss} S^{\kappa}_{s's'} \right. \label{B11115} \\
&&\hspace{-0.6cm} \left. +\pi \sum_{\bar{s}}
\left[ \tilde{\alpha}_{\kappa\kappa'}
(\epsilon,\omega_{s\bar{s}}) S^{\kappa}_{s\bar{s}}
S^{\kappa'}_{\bar{s}s} + \tilde{\alpha}_{\kappa'\kappa}
(\epsilon,\omega_{s'\bar{s}}) S^{\kappa'}_{s'\bar{s}}
S^{\kappa}_{\bar{s}s'} \right] \right\} \nonumber
\end{eqnarray}
and where the modified Bohr frequencies are given by
\begin{eqnarray}
\Omega_{ss'} &=& \omega_{ss'} - \frac{\lambda^2}{\hbar^2} \sum_{\kappa \kappa'}
\sum_{\bar{s}} \left[ \int_{-\infty}^{\infty} d\omega {\rm P}
\frac{\tilde{\alpha}_{\kappa\kappa'}(\epsilon,\omega)}
{\omega+\omega_{\bar{s}s}} S^{\kappa}_{s\bar{s}} S^{\kappa'}_{\bar{s}s}
\right. \nonumber \\
&&\hspace{1.5cm} \left. -\int_{-\infty}^{\infty} d\omega {\rm P}
\frac{\tilde{\alpha}_{\kappa\kappa'}(\epsilon,\omega)}
{\omega+\omega_{\bar{s}s'}} S^{\kappa}_{s'\bar{s}} S^{\kappa'}_{\bar{s}s'}
\right] \label{B11116} \; .
\end{eqnarray}
This equation is local in the energy of the environment.
%\begin{eqnarray}
%\dot{P}^I_{ss'}(\epsilon,t) &=&
%\frac{\lambda^2}{\hbar^2} \sum_{\kappa \kappa'} \left\{
%+ 2 \pi \tilde{\alpha}_{\kappa\kappa'}(\epsilon,0)
%S^{\kappa'}_{ss} S^{\kappa}_{s's'} \right. \label{B11114} \\
%&&\hspace{-1.7cm} -\pi \sum_{\bar{s}} \left( \tilde{\alpha}_{\kappa\kappa'}
%(\epsilon,\omega_{s\bar{s}}) S^{\kappa}_{s\bar{s}}
%S^{\kappa}_{\bar{s}s} + \tilde{\alpha}_{\kappa\kappa'}
%(\epsilon,\omega_{s'\bar{s}}) S^{\kappa}_{s'\bar{s}}
%S^{\kappa}_{\bar{s}s'} \right) \nonumber \\
%&&\hspace{-1.7cm}+i \sum_{\bar{s}} \left( \int_{-\infty}^{\infty}
%d\omega {\rm P}
%\frac{\tilde{\alpha}_{\kappa\kappa'}(\epsilon,\omega)}
%{\omega+\omega_{\bar{s}s}} S^{\kappa}_{s\bar{s}} S^{\kappa'}_{\bar{s}s}
%\right. \nonumber \\ &&\hspace{-0.3cm} \left. \left. -\int_{-\infty}^{\infty}
%d\omega {\rm P} \frac{\tilde{\alpha}_{\kappa\kappa'}(\epsilon,\omega)}
%{\omega+\omega_{\bar{s}s'}} S^{\kappa}_{s'\bar{s}} S^{\kappa'}_{\bar{s}s'}
%\right) \right\} P^I_{ss'}(\epsilon,t) \nonumber  \; ,
%\end{eqnarray}
This is not the case of the equation which rules the population dynamics
\begin{eqnarray}
\dot{P}_{ss}(\epsilon,t) &=&
\frac{\lambda^2}{\hbar^2} \sum_{\kappa \kappa'} \sum_{\bar{s}}
\Big\lbrace \label{B11117} \\
&&\hspace{-1.7cm}-2 \pi
\tilde{\alpha}_{\kappa\kappa'}
(\epsilon,-\omega_{\bar{s}s}) S^{\kappa}_{s\bar{s}} S^{\kappa'}_{\bar{s}s}
P_{ss}(\epsilon,t)
\nonumber \\ &&\hspace{-1.7cm}+2 \pi \tilde{\alpha}_{\kappa'\kappa}
(\epsilon-\hbar\omega_{\bar{s}s},\omega_{\bar{s}s}) S^{\kappa}_{s\bar{s}}
S^{\kappa'}_{\bar{s}s} P_{\bar{s}\bar{s}}(\epsilon-\hbar\omega_{\bar{s}s},t)
\Big\rbrace \nonumber \; .
\end{eqnarray}
This equation is a kind of Pauli equation for the total system
which preserves the unperturbed energy of the total system
\begin{eqnarray}
\mean{H_0}_t &=& \int d\epsilon \sum_s (E_s+\epsilon) P_{ss}(\epsilon,t)
\label{B11117E} \\ &=& \int d\varepsilon \; \varepsilon \; f(\varepsilon)
\; , \nonumber
\end{eqnarray}
where
\begin{eqnarray}
f(\varepsilon) = \sum_s P_{ss}(\varepsilon-E_s,t) \label{B11118}
\end{eqnarray}
represents the probability distribution inside a given energy shell
of energy $\varepsilon$ of the unperturbed total system.
Indeed, using Eq. (\ref{B11117}), we find
\begin{eqnarray}
\dot{f}(\varepsilon) = \sum_s \dot{P}_{ss}(\varepsilon-E_s,t) = 0 \;.
\label{B11119}
\end{eqnarray}
This shows that the energy of the total system is conserved by the
dynamics because the probability is preserved inside each energy shell
of the total system. If the initial condition of the total system is
a product of a subsystem pure state of energy $E_s$ with a microcanonical
distribution at energy $\epsilon_0$ for the environment
[$P(\epsilon,0)=\ket{s}\bra{s} \delta(\epsilon-\epsilon_0)$]
the energy distribution of the total system corresponds to
$f(\varepsilon) = \delta(\varepsilon-E_s-\epsilon_0)$.
If the subsystem is initially described by a density matrix
$\rho_S(0)$, we get $f(\varepsilon) = \sum_s \rho_{S,ss}(0)
\delta(\varepsilon-E_s-\epsilon_0)$. The population dynamics
independently evolves in each energy shell of the total system.
The equilibrium distribution of Eq. (\ref{B11117}) is
\begin{eqnarray}
\frac{P_{\bar{s}\bar{s}}(\epsilon-\hbar \omega_{\bar{s}s},\infty)}
{P_{ss}(\epsilon,\infty)} = \frac{\tilde{\alpha}_{\kappa\kappa'}
(\epsilon,-\omega_{\bar{s}s})}{\tilde{\alpha}_{\kappa'\kappa}
(\epsilon-\hbar\omega_{\bar{s}s},\omega_{\bar{s}s})} \; .
\label{B11120}
\end{eqnarray}
Using Eq. (\ref{A1112i}) and $\varepsilon=\epsilon+E_s$,
Eq. (\ref{B11120}) becomes
\begin{eqnarray}
\frac{P_{\bar{s}\bar{s}}(\varepsilon-E_{\bar{s}},\infty)}
{P_{ss}(\varepsilon-E_{s},\infty)} = \frac{n(\varepsilon-E_{\bar{s}})}
{n(\varepsilon-E_{s})} \; ,
\label{B11121}
\end{eqnarray}
which expresses detailed balance at equilibrium.
Because $f(\varepsilon)$ is invariant under the dynamics, we finally get
\begin{eqnarray}
P_{ss}(\varepsilon-E_{s},\infty)=\frac{n(\varepsilon-E_s)}{\sum_{\bar{s}}
n(\varepsilon-E_{\bar{s}})} f(\varepsilon) \; .
\label{B11122}
\end{eqnarray}
At equilibrium, each quantum level inside a given energy shell
of the total system has the same probability.
This means that our equation describes how any initial
distribution inside a given energy shell of the total
system reaches equilibrium.

%%%%%%%%%%%%%%%%%%%%%%%%%%%%%%%%%%%%%%%%%%%%%%%%%%%%%%%%%%%%%%%%%%%%
\section{Equivalence between the ensembles}\label{equivens}

In this section, we start by showing using a simple
qualitative argument that if the environment is initially
described by a canonical distribution and is assumed large
enough for not being affected by the subsystem dynamics, our
equation reduces to the Redfield master equation.
However, the most important part of this section is devoted to the problem of
understanding in detail how our master equation (which
rules the dynamics of a subsystem interacting with
an environment described by a microcanonical distribution)
effectively reduces to a Redfield master equation
(which rules the dynamics of a subsystem interacting
with an environment described by a canonical distribution)
if the equivalence between the microcanonical and the canonical
ensemble is satisfied for the environment.

The canonical density matrix of the environment is related
to the microcanonical density matrix of the environment by
\begin{eqnarray}
\rho^{eq}_{can}(\beta)=\frac{\e^{-\beta H_B}}{Z} = \int d\epsilon
\; W(\epsilon) \; \rho^{eq}_{mic}(\epsilon) \; ,
\label{C11111}
\end{eqnarray}
where
\begin{eqnarray}
W(\epsilon) = \frac{\e^{-\beta \epsilon} n(\epsilon)}{Z} \; .
\label{C11112}
\end{eqnarray}
We notice that the condition (\ref{A1112i}) found for the
microcanonical correlation functions is in fact at the origin
of the KMS condition for the canonical correlation functions.
Indeed, using (\ref{A1112i}) and
\begin{eqnarray}
\tilde{\alpha}_{\kappa\kappa'}(\beta,\omega)= \int d\epsilon
\; W(\epsilon) \; \tilde{\alpha}_{\kappa\kappa'}(\epsilon,\omega) \; ,
\label{C11113}
\end{eqnarray}
we obtain the KMS condition
\begin{eqnarray}
\tilde{\alpha}_{\kappa\kappa'}(\beta,\omega) =
\e^{\beta \hbar \omega}
\tilde{\alpha}_{\kappa'\kappa}(\beta,-\omega) \; .
\label{C11114}
\end{eqnarray}
A very easy way to see the link between our equation and
the Redfield equation is to assume that the environment
is initially described by a canonical distribution and remains
in it at any time during its interaction with the subsystem.
This assumption can be expressed by the ansatz
\begin{eqnarray}
P(\epsilon,t) \approx \rho_S(t)
\frac{\e^{-\beta \epsilon}n(\epsilon)}{Z} \; .
\label{C11115}
\end{eqnarray}
This can be qualitatively justified by assuming that
the environment is very large compared to the
subsystem. By integrating Eq. (\ref{A1112j}) over the energy
of the environment and with the help of Eq. (\ref{A1112d}),
Eq. (\ref{C11113}) and (\ref{C11115}), we get the Redfield equation
\cite{Red57,KampenB97,KuboB98b,GaspNaga99,Breuer02}
\begin{eqnarray}
\dot{\rho}_S(t)&=& {\cal L}_S \rho_S(t) + \frac{\lambda^2}{\hbar^2}
\sum_{\kappa\kappa'} \int_{0}^{t} d\tau \Big\{ \label{C11116} \\
&&\hspace{-1.5cm}-\alpha_{\kappa \kappa'}(\beta,\tau) S^{\kappa}
S^{\kappa'}(-\tau) \rho_S(t) - \alpha^{*}_{\kappa \kappa'}(\beta,\tau)
\rho_S(t) S^{\kappa'}(-\tau) S^{\kappa} \nonumber \\
&&\hspace{-1.5cm}+\alpha^{*}_{\kappa \kappa'}(\beta,\tau) S^{\kappa}
\rho_S(t) S^{\kappa'}(-\tau) + \alpha_{\kappa \kappa'}(\beta,\tau)
S^{\kappa'}(-\tau) \rho_S(t) S^{\kappa} \Big\} \; . \nonumber
\end{eqnarray}
with $\alpha(\tau)=\int d\omega \e^{-i\omega \tau}
\alpha(\omega)$.
Exactly the same procedure can be used for Eq. (\ref{B11113})
and we thus get the RWA version of the Redfield equation
(also called the weak-coupling-limit master equation)
\cite{Breuer02,Spohn78,Spohn80,CohenTann96,Gardiner00}.
In this last case, one can also show that, by integrating the
equilibrium distribution (\ref{B11120}) over the energy of the
environment and by using (\ref{C11115}) and the KMS condition
(\ref{C11114}), we get
\begin{eqnarray}
\frac{\rho^{S}_{\bar{s}\bar{s}}}{\rho^{S}_{ss}}=
\e^{-\beta \hbar \omega_{\bar{s}s}} \; .
\label{C11117}
\end{eqnarray}
As expected, using the normalization condition $\sum_{\bar{s}}
\rho^{S}_{\bar{s}\bar{s}}=1$, we find that the subsystem
equilibrium distribution of the RWA form of the Redfield equation
is a canonical distribution at the temperature of the environment
\begin{eqnarray}
\rho^{S}_{ss} = \frac{\e^{-\beta E_s}}{Z_S} \; ,
\label{C11118}
\end{eqnarray}
where $Z_S=\sum_s \e^{-\beta E_s}$.

After this qualitative discussion, we now show that the
precise condition for the Redfield equation to provide an effective
description of our master equation is the equivalence between
the canonical and microcanonical ensembles.
Integrating Eq. (\ref{A1112j}) over the energy of
the environment and using (\ref{A1112d}), we get
\begin{eqnarray}
\dot{\rho}_S(t) &=& {\cal L}_S \rho_S(t)
+ \frac{\lambda^2}{\hbar^2} \sum_{\kappa \kappa'}
\int_{0}^{t} d\tau \int_{-\infty}^{\infty} d\omega
\Big\lbrace \label{C11119} \\
&&\hspace{-1cm}-\e^{-i \omega \tau} \int d\epsilon \;
\tilde{\alpha}_{\kappa\kappa'}(\epsilon,\omega)
S^{\kappa} S^{\kappa'}(-\tau) P(\epsilon,t) \nonumber \\
&&\hspace{-1cm}-\e^{i \omega \tau} \int d\epsilon \;
\tilde{\alpha}_{\kappa'\kappa}(\epsilon,\omega)
P(\epsilon,t) S^{\kappa'}(-\tau) S^{\kappa} \nonumber \\
&&\hspace{-1cm}+\e^{-i \omega \tau} \int d\epsilon \;
\tilde{\alpha}_{\kappa'\kappa}(\epsilon,-\omega)
S^{\kappa} P(\epsilon,t) S^{\kappa'}(-\tau) \nonumber \\
&&\hspace{-1cm}+\e^{i \omega \tau} \int d\epsilon \;
\tilde{\alpha}_{\kappa\kappa'}(\epsilon,-\omega)
S^{\kappa'}(-\tau) P(\epsilon,t) S^{\kappa} \Big\rbrace \; .
\nonumber
\end{eqnarray}
In order to close the equation for the subsystem density
matrix, we have to assume that the microcanonical correlation
functions can be put out of the energy integral.
This can only be justified if the environment is such that
\begin{eqnarray}
\tilde{\alpha}_{\kappa\kappa'}(\epsilon+\hbar \omega_S,\omega)
\approx \tilde{\alpha}_{\kappa\kappa'}(\epsilon,\omega) \; .
\label{C11120}
\end{eqnarray}
By doing this, we obtain Eq. (\ref{C11116}), but where the
canonical correlation functions have to be replaced by
the microcanonical correlation function
$\tilde{\alpha}_{\kappa \kappa'}(\beta,\omega) \to
\tilde{\alpha}_{\kappa \kappa'}(\epsilon,\omega)$.
Therefore, in order to obtain the Redfield
equation, we need to further assume that we can identify
the microcanonical with the canonical correlation
functions of the environment
\begin{eqnarray}
\tilde{\alpha}_{\kappa \kappa'}(\epsilon,\omega) \approx
\tilde{\alpha}_{\kappa \kappa'}(\beta,\omega) \; .
\label{C11121}
\end{eqnarray}
Let us now find the conditions under which the two
assumptions (\ref{C11120}) and (\ref{C11121}) are valid.
The entropy, associated with the microcanonical distribution
in an energy shell of the environment of width
$\delta_{\epsilon}$ and corresponding to an energy $\epsilon$,
is given by
\begin{eqnarray}
S(\epsilon) \equiv k_B \ln w
= k_B \ln n(\epsilon) \delta_{\epsilon} \; ,
\label{C11122}
\end{eqnarray}
where $w$ is the complexion number (i.e., the number of available
states in the energy shell). The microcanonical temperature
of the environment associated with this microcanonical entropy
is defined as
\begin{eqnarray}
\beta(\epsilon)=\frac{1}{k_B T(\epsilon)} \equiv
\frac{1}{k_B} \frac{dS(\epsilon)}{d\epsilon}
= \frac{1}{n(\epsilon)} \frac{dn(\epsilon)}{d\epsilon} \; .
\label{C11123}
\end{eqnarray}
Suppose that the environment is in a microcanonical distribution at
the energy $\epsilon_m$. This environment can be effectively
described by a canonical distribution at the temperature $(k_B\beta)^{-1}$
if $W(\epsilon)$ is a sharp function with its maximum at the energy
$\epsilon_m$, which is therefore the most probable energy.
In this case, we can expand $\ln W(\epsilon)$ around $\epsilon_m$.
We get
\begin{eqnarray}
\ln W(\epsilon) = \ln W(\epsilon_m) + \left\lbrack
\frac{d^2}{d\epsilon^2} \ln W(\epsilon)
\right\rbrack_{\epsilon=\epsilon_m} \frac{(\epsilon-\epsilon_m)^2}{2!}
\nonumber \\ + \left\lbrack \frac{d^3}{d\epsilon^3} \ln W(\epsilon)
\right\rbrack_{\epsilon=\epsilon_m} \frac{(\epsilon-\epsilon_m)^3}{3!}
+ \hdots \; ,
\label{C11124}
\end{eqnarray}
since $d\ln W(\epsilon_m)/d\epsilon=0$.
Because the energy $\epsilon_m$ corresponds to the maximum
of $\ln W(\epsilon)$, the canonical temperature is
equal to the microcanonical temperature at $\epsilon_m$
\begin{eqnarray}
\beta = \beta(\epsilon_m) \; .
\label{C11125}
\end{eqnarray}
Now using the microcanonical heat capacity defined by
\begin{eqnarray}
\frac{1}{C(\epsilon)} \equiv \frac{d T(\epsilon)}{d\epsilon} \; ,
\label{C11126}
\end{eqnarray}
we have that
\begin{eqnarray}
\frac{d^2}{d\epsilon^2} \ln W(\epsilon) =
\frac{d \beta(\epsilon)}{d\epsilon} =
-\frac{1}{k_B T^2(\epsilon) C(\epsilon)} \; .
\label{C11127}
\end{eqnarray}
We see now that the rational to truncate the series
(\ref{C11124}) is a large and positive heat capacity.
This is true for most environments in the limit of a
large number of degrees of freedom $N$, since typically
$C(\epsilon) \sim N$.
We can now rewrite Eq. (\ref{C11124}) as
\begin{eqnarray}
W(\epsilon) \approx W(\epsilon_m)
\exp{\left[ -\frac{(\epsilon-\epsilon_m)^2}{2\sigma^2(\epsilon_m)}
\right]} \; , \label{C11128}
\end{eqnarray}
where
\begin{eqnarray}
\sigma^2(\epsilon_m) = k_B T^2(\epsilon_m) C(\epsilon_m) \; .
\label{C11129}
\end{eqnarray}
This result confirms that $\epsilon_m$ is the maximum of
$W(\epsilon)$, and also shows that $\epsilon_m$ is its
mean value if $\sigma(\epsilon_m)$ is small compared
to the typical energies of the environment.
This is true for large $N$, since
$\epsilon/\sigma(\epsilon_m) \sim \sqrt{N}$.
Under these conditions, an environment described by a microcanonical
distribution at the energy $\epsilon_m$ can be effectively described
by a canonical distribution at the temperature $\beta(\epsilon_m)$ so
that the second assumption (\ref{C11121}) becomes justified.
However, in order to justify the first assumption (\ref{C11120}),
the environment effective canonical temperature has also to remain
unchanged under energy shifts of the environment energy of the
order of the typical subsystem quanta $\hbar \omega_S$
\begin{eqnarray}
\beta(\epsilon \pm \hbar \omega_S) = \beta(\epsilon) \pm
\frac{d \beta(\epsilon)}{d\epsilon} \, \hbar \omega_S
+ \hdots \; .
\label{C11130}
\end{eqnarray}
The condition to have such an isothermal environment is
\begin{eqnarray}
\Big\vert \frac{\hbar \omega_S}{\beta(\epsilon)}
\frac{d\beta(\epsilon)}{d\epsilon} \Big\vert =
\Big\vert \frac{\hbar \omega_S}{T(\epsilon)C(\epsilon)}
\Big\vert \ll 1 \; . \label{C11131}
\end{eqnarray}
This condition is again satisfied when the number of
degrees of freedom of the environment becomes large
because $\hbar \omega_S/(T(\epsilon)C(\epsilon)) \sim 1/N$.
Our conclusion is that, in the limit of an infinitely large
environment $N \to \infty$, $\sigma(\epsilon)$ becomes
infinitely small compared to typical environment energy
scales (so that we have the equivalence between the microcanonical
and canonical ensembles) but also becomes infinitely large
compared to typical subsystem energy scales $\hbar \omega_S$
(so that the environment is isothermal for the subsystem).
It is in this limit that the Redfield master equation becomes
valid and can provide an effective description of our
master equation.

%%%%%%%%%%%%%%%%%%%%%%%%%%%%%%%%%%%%%%%%%%%%%%%%%%%%%%%%%%%%%%%%%%%%%%%%%%%%%%%%%%%%%%%%%%%%%%%%%%%%%
\section{Conclusions \label{conclusion}}

In this paper, we have considered a subsystem interacting with
an environment which has a sufficiently large number of degrees
of freedom so that its spectrum can be supposed quasi-continuous.
As a consequence, this environment can drive the subsystem into
a relaxation process on time scales typically shorter than the
Heisenberg time of the environment [$t_H = \hbar n(\epsilon)$, where
$n(\epsilon)$ is the average density of states of the environment].
However, the number of degrees of freedom of this environment
may still be too small for the transitions to leave it isothermal.
Indeed, the heat capacity is finite and the energy exchanges
between the subsystem and the environment can modify
the energy distribution of the environment.
This is not an unrealistic assumption since the density of states
of the environment grows exponentially with the number of degrees
of freedom albeit the heat capacity only grows linearly.
By using projection operators, we derived a master equation
governing the relaxation dynamics of such a subsystem.
This equation describes the evolution of the subsystem
density matrix distributed over the energy of the environment.
This allows us to take into account the changes of the environment
energy distribution due to its interaction with the subsystem.
By performing the rotating wave approximation (RWA) on this
master equation, we have been able to show that the subsystem
populations evolve independently from the coherences.
The coherences decay in the form of exponentially damped
oscillations, while the populations obey a kind of Pauli
equation for the total system.
This equation conserves the energy of the total system and its
equilibrium distribution corresponds to the uniform distribution of
the probability over the energy shell of the unperturbed total system.
Our equation provides a natural representation of the dynamics of a
subsystem interacting with an environment described by a
microcanonical distribution.
The Redfield master equation is the usual way to represent
the dynamics of a subsystem interacting with an environment
described by a canonical distribution and is a closed equation
for the density matrix of the subsystem.
We have shown that, if the equivalence between the microcanonical
and canonical ensembles is satisfied for the environment
(containing infinitely many degrees of freedom), our equation
reduces to the Redfield master equation.
If this equivalence is not satisfied, our equation becomes of crucial
importance to correctly describe the subsystem relaxation.
In the same sense that the microcanonical ensemble is more
fundamental than the canonical ensemble, our master equation
is more fundamental then the Redfield equations.
We believe that this work improves the understanding of kinetic
processes in nanosystems where the thermodynamical
limit cannot always be taken.

%%%%%%%%%%%%%%%%%%%%%%%%%%%%%%%%%%%%%%%%%%%%%%%%%%%%%%%%%%%%%%%%%%%%%%%%%%%%%%%%%%%%%%%%%%%%%%%%%%%%%
%%%%%%%%%%%%%%%%%%%%%%%%%%%%%%%%%%%%%%%%%%%%%%%%%%%%%%%%%%%%%%%%%%%%%%%%%%%%%%%%%%%%%%%%%%%%%%%%%%%%%

\begin{acknowledgments}
M. E. thanks the F.R.S.-FNRS Belgium for financial support.
This research is financially supported by the
"Communaut\'e fran\c caise de Belgique" (contract "Actions de
Recherche Concert\'ees" No. 04/09-312) and the F.R.S.-FNRS
Belgium (contract F.R.F.C. Nos. 2.4542.02 and 2.4577.04).
\end{acknowledgments}

%%%%%%%%%%%%%%%%%%%%%%%%%%%%%%%%%%%%%%%%%%%%%%%%%%%%%%%%%%%%%%%%%%%%%%%%%%%%%%%%%%%%%%%%%%%%%%%%%%%%%


\begin{thebibliography} {1}

\bibitem{EspoGaspPRE03a}
M. Esposito and P. Gaspard, Phys. Rev. E {\bf 68}, 066112 (2003).

\bibitem{EspositoPRE03b}
M. Esposito and P. Gaspard, Phys. Rev. E {\bf 68}, 066113 (2003).

\bibitem{KuboB98a}
R. Kubo, M. Toda, and N. Saito, Statistical Physics I: Equilibrium
Statistical Mechanics, 2nd ed. (Springer, Berlin, 1998).

\bibitem{Red57}
A. G. Redfield, IBM J. Res. Dev. {\bf 1}, 19 (1957).

\bibitem{KampenB97}
N. G. van Kampen, Stochastic Processes in Physics and
Chemistry, 2nd ed. (North-Holland, Amsterdam, 1997).

\bibitem{KuboB98b}
R. Kubo, M. Toda, and N. Hashitsume, Statistical Physics II:
Nonequilibrium Statistical Mechanics, 2nd ed. (Springer,
Berlin, 1998).

\bibitem{GaspNaga99}
P. Gaspard and M. Nagaoka, J. Chem. Phys. {\bf 111}, 5668 (1999).

\bibitem{Breuer02}
H.-P. Breuer and F. Petruccione, The Theory of Open Quantum
Systems (Oxford University Press, Oxford, 2002).

\bibitem{Haake73}
F. Haake, Statistical Treatment of Open Systems,
Springer Tracts in Modern Physics, Vol. 66 (Springer, Berlin, 1973).

\bibitem{Spohn78}
H. Spohn and J. L. Lebowitz, Adv. Chem. Phys. {\bf 38}, 109 (1978).

\bibitem{Spohn80}
H. Spohn, Rev. Mod. Phys. {\bf 53}, 569 (1980).

\bibitem{CohenTann96}
C. Cohen-Tannoudji, J. Dupont-Roc, and G. Grynberg, Processus
d'Interaction entre Photons et Atomes (CNRS Editions, Paris, 1996).

\bibitem{Gardiner00}
C. W. Gardiner and P. Zoller, Quantum Noise (Springer, Berlin, 2000).

\bibitem{Jensen85}
%Statistical Behavior in Deterministic Quantum Systems with Few 
%Degrees of Freedom
R. V. Jensen and R. Shankar , Phys. Rev. Lett. {\bf 54}, 1879 (1985).

\bibitem{Gross00}
D. H. E. Gross, Microcanonical Thermodynamics: Phase Transitions
in ``Small" Systems (World Scientific, Singapore, 2000).

\bibitem{Jellinek00}
%On the temperature, equipartition, degrees of freedom, and finite
%size effects: Application to aluminum clusters
J. Jellinek and A. Goldberg, J. Chem. Phys. {\bf 113}, 2570 (2000).

\bibitem{Shirts02}
%Deviations from the Boltzmann distribution in small microcanonical quantum
%systems: Two approximate one-particle energy distributions
R. B. Shirts and M. R. Shirts, J. Chem. Phys. {\bf 117}, 5564 (2002).

\bibitem{Gross05}
%The microcanonical thermodynamics of finite systems:
%The microscopic origin of condensation and phase separations,
%and the conditions for heat flow from lower to higher temperatures
D. H. E. Gross and J. F. Kenney, J. Chem. Phys. {\bf 122}, 224111 (2005).

\bibitem{Janke06}
%Microcanonical Analyses of Peptide Aggregation Processes
C. Junghans, M. Bachmann, and W. Janke, Phys. Rev. Lett. {\bf 97}, 
218103 (2006).

\bibitem{Barre01}
%Inequivalence of Ensembles in a System with Long-Range Interactions
J. Barr\'e, D. Mukamel, and S. Ruffo, Phys. Rev. Lett. {\bf 87}, 030601 (2001).

\bibitem{Cohen01}
%On first-order phase transitions in microcanonical and canonical 
non-extensive systems
I. Ispolatov and E. G. D. Cohen, Physica A {\bf 295}, 475 (2001).

\bibitem{Shepelyansky01}
%Quantum chaos and quantum computers
%D. L. Shepelyansky, Phys. Scr. T90 {\bf T90}, 112 (2001);
%Emergence of Fermi-Dirac thermalization in the quantum computer core
G. Benenti, G. Casati, and D.L. Shepelyansky, Eur. Phys. J. D {\bf 
17}, 265 (2001).

\bibitem{Barre02}
%Microcanonical solution of lattice models with long-range interactions.
J. Barr\'e, Physica A {\bf 305}, 172 (2002).

\bibitem{Thirring71}
%A Soluble Model for a System with Negative Specific Heat
P. Hertel and W. Thirring, Annals of Physics {\bf 63}, 520 (1971).

\bibitem{Gross97}
%Microcanonical thermodynamics and statistical fragmentation
%of dissipative systems. The topological structure of the N-body space
D. H. E. Gross, Phys. Rep. {\bf 279}, 119 (1997).

\bibitem{Ruffo02}
%First- and second-order clustering transitions for a system with
%infinite-range attractive interaction
M. Antoni, S. Ruffo, and A. Torcini, Phys. Rev. E {\bf 66}, 025103(R) (2002).

\bibitem{Touchette04}
%An Introduction to the Thermodynamic and Macrostate
%Levels of Nonequivalent Ensembles
H. Touchette, R. S. Ellis and B. Turkington, Physica A {\bf 340}, 138 (2004).

\bibitem{Mukamel95}
S. Mukamel, Principles of Nonlinear Optical Spectroscopy
(Oxford University Press, New York, 1995).

\bibitem{Lindblad76}
G. Lindblad, Commun. Math. Phys. {\bf 48}, 119 (1976).

\bibitem{Espositothese}
M. Esposito, cond-mat/0412495.

\bibitem{Budini}
%Lindblad rate equations
A. A. Budini, Phys. Rev. A {\bf 74} 053815 (2006).

\bibitem{BreuerGemmer}
H. P. Breuer, J. Gemmer and M. Michel, Phys. Rev. E {\bf 73}, 016139 (2006).

\bibitem{Breuer}
%Non-Markovian generalization of the Lindblad theory of open quantum systems
H. P. Breuer, Phys. Rev. A {\bf 75}, 022103 (2007).

\bibitem{Silbey92}
A. Suarez, R. Silbey and I. Oppenheim, J. Chem. Phys. {\bf 97} 5101 (1992).

\bibitem{Silbey05}
%Markovian Approximation in the Relaxation of Open Quantum Systems
Y. C. Cheng and R. J. Silbey, J. Phys. Chem. B {\bf 109}, 21399 (2005).

\bibitem{Tannor97}
%Phase space approach to theories of quantum dissipation
D. Kohen, C. C. Marston, and D. J. Tannor, J. Chem. Phys. {\bf 107} 
5236 (1997).

%\bibitem{localheating}
%Nonlinear Annihilation of Excitations in Photosynthetic Systems
%L. Valkunas, G. Trinkunas, V. Liuolia, and R. van Grondellet,
%Biophysical Journal {\bf 69}, 1117 (1995);
%Singlet-Singlet Annihilation and Local Heating in FMO Complexes
%V. Gulbinas, L. Valkunas, D. Kuciauskas, E. Katilius,
%V. Liuolia, W. Zhou, and R. E. Blankenship,
%J. Phys. Chem. {\bf 100}, 17950 (1996).

\end{thebibliography}
\end{document}